\documentclass[10pt, twocolumn]{IEEEtran/IEEEtran}
\usepackage{cite}
\usepackage{graphicx}
\usepackage{url}
\usepackage{epsfig,color}
\usepackage[centertags]{amsmath}
\usepackage{algorithm}
\usepackage{algorithmic}
\usepackage{amsfonts}
\newcommand{\beq}{\begin{equation}}
\newcommand{\eeq}{\end{equation}}
\newcommand{\beqn}{\begin{align}}
\newcommand{\eeqn}{\end{align}}

\begin{document}

\title{Low-Complexity QoS-Aware Coordinated Scheduling for Heterogenous Networks}

\author{Jun~Zhu,~\IEEEmembership{Student Member,~IEEE} and
Hong-Chuan~Yang,~\IEEEmembership{Senior Member,~IEEE}}
	
%\author{Jun~Zhu and Hong-Chuan~Yang,~\IEEEmembership{Senior Member,~IEEE}

%\thanks{This work was supported by a Discovery Grant from NSERC.}}

\IEEEoverridecommandlockouts

\setcounter{page}{1}

\maketitle

\begin{abstract}
In this paper, we consider a heterogenous network (HetNet), where low-power indoor femtocells are deployed in the coverage area of the existing macro base station (MBS). This paper proposes a novel coordinated random beamforming and user scheduling strategy to improve the throughput of users served by the femtocell access point (FAP) while satisfying the quality-of-service (QoS) requirements of users served by both MBS and FAP. The strategy, termed as QoS-Aware Coodinated Scheduling (QACS), requires limited coordination between the MBS and FAP, i.e., only the indexes of the qualified beams are shared. Exact statistical analysis for the ergodic achievable rate of both FAP and MBS with the proposed strategy are presented. Scheduling fairness is also addressed for the proposed QACS.
\end{abstract}
%\vspace{-6mm}
%\begin{keywords}
%Heterogenous network, quality of service, coodinated scheduling, interference mitigation, random unitary beamforming, and performance analysis.
%\end{keywords}
\vspace{-6mm}
\section{Introduction}
In order to meet the unprecedented demands for higher data rates in wireless networks, the concept of heterogeneous network (HetNet) has been proposed for the next generation wireless systems \cite{hetnet}. HetNet is able to increase network throughput by deploying smaller cells within the coverage of existing macro cellular base station (BS). One of the successful sample application scenarios is macro-femto HetNet, which has drawn tremendous attentions from academia, industry, and business, and has already been included in the LTE-advanced standardization process \cite{femto1,femto2,femto3}. A femtocell access point (FAP) is a low-power, low-cost wireless access point that typically operates indoor over a licensed spectrum to provide short-range and high-speed service. These FAPs are connected to the core service network via residential Digital Subscriber Line (DSL) or broadband cables as their backhaul channel \cite{femto3}. The development of femtocell has alleviated the poor indoor coverage of the conventional macro BS (MBS). This is of great significance as it has been shown that over 60$\%$ of cellular voice calls and over 90$\%$ of cellular data services are requested by indoor subscribers \cite{Ding}.

The main challenge of the macro-femto HetNet lies in the cross-tier interference management, required by the spectrum sharing in the HetNet \cite{interference}. Cross-tier interference can seriously limit the throughput gain in both macrocell and femtocell, and thus, it is essential to develop effective interference mitigation and/or cancelation techniques for such two-tier networks \cite{Ding,icc_zhu,dai}. On the other hand, multiple-input multiple-output (MIMO) techniques can provide order-of-magnitude improvements of the spectral efficiency of wireless systems, and thus has already widely been adopted in current commercial systems \cite{mimohetnet}. Therefore, one possible approach to solve the interference management problem in the literature is referred to as coordinated multiple point (CoMP) transmission with joint precoding/transmission \cite{comp,LG09}. Although the concept of CoMP is able to improve cell coverage as well as the overall system throughput, it usually requires a large amount of overhead signaling, which limits their applicability in real-world systems. Note that the wired connections between BSs, usually through the mobile switching center, are already fully loaded with the increasing amount of multimedia data traffic.

Coordinated scheduling and beamforming techniques represent more practical solutions for interference mitigation in HetNets or even a general multi-cell network \cite{huang,gc_zhu,eurasip_zhu,gc2_zhu}, as they require no data and limited CSI sharing. The good/bad  precoding matrix index (PMI) reporting algorithm proposed for LTE-advanced is the best example of this category \cite{3GPP, LG09a}, where only the preferred or the restricted PMI needs to be exchanged between MBSs. Similar methods have been considered in the macro-femto setup in e.g. \cite{Ding,icc_zhu,dai}. In particular, \cite{icc_zhu} and \cite{dai} proposed coordinated scheduling and beamforming schemes, where only limited information are shared between the MBS and the FAP, in order to alleviate the burden of backhaul connections. However, neither of them takes into account the qualify-of-service (QoS) requirements of both macrocell and femtocell users. This becomes more important when wireless systems evolve towards the fifth generation (5G) \cite{hetnet5G}, which should efficiently support various levels of QoS requirements from diverse types of wireless applications and services. A family of cross-tier interference mitigation approaches has been developed in \cite{Ding} based on MIMO precoding, in which, the QoS requirements of macrocell/femtocell users are considered, but with the absence of multiuser scheduling. All results presented in \cite{dai} are obtained from Monte-Carlo simulations, while in this work, we perform exact statistical analysis to acquire analytical results for the ergodic rate achieved by both FAP and MBS. We can then investigate the relationship between system performance and various parameters based on such results. Moreover, statistical analysis upon QoS-aware scheduing are rarely investigated due to its higher complexity, compared with conventional coordinated beamforming reported in \cite{Ding} and \cite{icc_zhu}.

In this paper, we propose and analytically evaluate the performance of a low-complexity coordinated beamforming and scheduling scheme for a macro-femto HetNet based on random unitary beamforming transmission \cite{Hassibi}, named as QoS-Aware Coordinated Scheduling (QACS). The proposed QACS manages to suppress the interference from macrocell to femtocell while the QoS requirements of both macrocell and femtocell users are guaranteed. In particular, a subset of beams are allowed to be adopted at the macrocell based upon the amount of interference that they generate to the selected user in femtocell. QACS involves limited overhead signaling between MBS and FAP, as only several indexes of beams need to be shared. Moreover, unlike the work in \cite{icc_zhu}, where the user {{near}} MBS/FAP always has the priority to be selected in order to improve the overall system throughput, we consider the fairness in scheduling among all users in the macrocell/femtocell. Unlike simulation based investigation in previous work, e.g., \cite{dai}, we address the exact throughput analysis of the resulting coordinated two-tier network under the proposed QACS. In particular, we offer exact expressions for the ergodic throughput when the MBS/FAP is serving one selected user in its own coverage. Based on the expressions, we then can study the relation between throughput and various system parameters including the number of transmit antennas, the number of users, the feedback load, and the QoS requirements. The proposed scheme and the associated analytical results will provide important design guidelines for future cellular systems.

%The remainder of the paper is organized as follows. In Section II, we outline the system and
%channel models under consideration. Section III presents our proposed coordinated multiple beam and user selection strategy. The throughput analysis of the proposed
%system under QACS are given in Section IV. Section VI provides selected analytical and simulation results. The paper concludes in Section VII.
%
%\textit{Notations:} Symbols ${\bf X}^T$ and ${\bf X}^H$ denote the transpose and Hermitian transpose of the matrix ${\bf X}$, respectively. $\|{\bf x}\|^2$ is the Euclidean norm square of the vector ${\bf x}$. $\mathbb{C}^{m\times n}$ represents the space of $m \times n$ matrices with complex-values elements.
\vspace{-6mm}
\section{System and Channel Models}\label{s2}
\begin{figure}
\centering
  \includegraphics[width=3in]{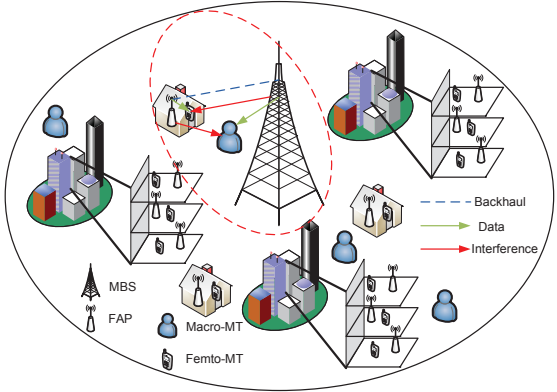}\vspace*{-4mm}\\
  \caption{{System model.}}\label{sys_mod}
  \end{figure}
In this section, we introduce the system and channel models for the considered two-tier heterogenous MIMO system, {{as depicted in Fig.~\ref{sys_mod}.}} For convenience, the notations used in this paper are defined in Table \ref{table1}.
%\begin{figure}
%\centering
%\includegraphics[width=2.5in]{sys_mod.eps}\\
%    \caption{A heterogeneous network model.}\label{sys_mod}
%\end{figure}
\vspace{-4mm}
\subsection{Two-Tier Heterogenous MIMO System Model}
\begin{table*}
\caption{Summary of most important variables used in this paper.}
 \label{table1}
   \centering
   \begin{tabular}{|l|l|}
         \hline
         \multicolumn{1}{|l|}{\bf{Symbols}} & \bf{Description}\\
         \hline
         $N_F (N_M), K_F (K_M)$ & Number of FAP (MBS) antennas, number of femto-MTs (macro-MTs) \\
         $\gamma_F (\gamma_M)$ & Received SINR of the selected femto-MT (macro-MT)\\
         $\Gamma_F (\Gamma_M)$ & SINR requirement of the selected femto-MT (macro-MT)\\
         $N_Q$ & Number of qualified beams in MBS's codebook such that $\gamma_F \geq \Gamma_F$\\
         $K_Q$ & Number of qualified macro-MTs such that $\gamma_M \geq \Gamma_M$\\
         $N_B$ & Number of best beams requested by qualified macro-MTs\\
         $P_F (P_M)$ & Transmit power for FAP (MBS)\\
         $\beta_{F,k} (\beta_{M,k}), \alpha_{F,k} (\alpha_{M,k})$ & Path-loss from FAP (MBS) to femto-MT $k$, from FAP (MBS) to macro-MT\\
         ${\bf h}_{F,k} ({\bf h}_{M,k}), {\bf g}_{F,k} ({\bf g}_{M,k})$ & Small-scale fading component from FAP (MBS) to femto-MT $k$, from FAP (MBS) to macro-MT $k$\\
         ${\bf w}_i ({\bf f}_j),\sigma_F^2 (\sigma_M^2)$ & Beam $i$ ($j$) for FAP (MBS),Gaussian noise at femto-MT (macro-MT)\\
         $\lambda_F (\mu_F),\lambda_{M,k} (\mu_{M,k})$ & $\lambda_F=\frac{P_M \beta_{M,k^*}}{P_F \beta_{F,k^*}} \left(\mu_F=\frac{\sigma^2_F}{P_F \beta_{F,k^*}}\right)$, $\lambda_{M,k}=\frac{P_M \alpha_{F,k}}{P_M \beta_{M,k}} \left(\mu_{M,k}=\frac{\sigma^2_M}{P_M \alpha_{M,k}}\right)$\\
           \hline
     \end{tabular}
 \end{table*}
The system under consideration consists one $N_F$-antenna FAP and one $N_M$-antenna MBS with overlapping coverage \footnote{{When deployed in practice, usually a number of femtocells are equipped in the coverage of one macrocell. Owing to the lower-power FAP and indoor operating environment, however, the selected macro-MT is only affected by its closest FAP, and the selected femto-MT is only affected by the MBS. Hence, our model, although simple, is of practical interests.}}. They work on the same radio spectrum to serve their respective scheduled mobile terminals (MTs). Each MT is equipped with a single-antenna due to size, cost, or complexity constraints, and there are $K_F$ femto-MTs and $K_M$ macro-MTs. The FAP is connected to its associated MBS via optical fiber backhaul links. The channel is assumed to be time invariant during one frame transmission, but time varying from one frame to another.
\vspace{-4mm}
\subsection{Channel Models}
Both the MBS and the FAP employ a codebook-based random unitary beamforming strategy to serve one selected user in their respective coverage area in each frame, in order to reduce the multiuser interference. In particular, the femtocell codebook has $N_F$ ortho-normal beams $\{{\bf w}_i\}^{N_F}_{i=1} \in \mathbb{C}^{N_F \times 1}$, while the macrocell codebook has $N_M$ ortho-normal beams $\{{\bf f}_j\}^{N_M}_{j=1} \in \mathbb{C}^{N_M \times 1}$, all randomly generated from isotropic distributions \cite{Hassibi}. The MBS/FAP then selects one of the beams from the codebook (as per the rules to be detailed below) to serve a selected user. Suppose both the beam and user are selected, the received symbol at the selected femto-MT, $y_{F}$ and that at the selected macro-MT, $y_{M}$ are obtained as
\begin{subequations}\label{y}
\begin{equation}
{{y_{F}=}}\sqrt{P_F \beta_{F,k^*}} {\bf h}^H_{F,k^*} {\bf w}_{i^*} s_{F,k^*}+ \sqrt{P_M \beta_{M,k^*}} {\bf h}^H_{M,k^*} {\bf f}_{j^*} s_{M,k^*}+n_F
\end{equation}
and
\begin{equation}
{{y_{M}=}}\sqrt{P_M \alpha_{M,k^*}} {\bf g}^H_{M,k^*} {\bf f}_{j^*} s_{M,k^*}+ \sqrt{P_F \alpha_{F,k^*}} {\bf g}^H_{F,k^*} {\bf w}_{i^*} s_{F,k^*}+n_M,
\end{equation}
\end{subequations}
respectively, where $s_{F,k^*}$ and $s_{M,k^*}$ are the data symbols intended for the selected femto-MT and macro-MT. ${\bf w}_{i^*}$ and ${\bf f}_{j^*}$ are the corresponding beams for femtocell and macrocell. Besides, $\beta_{F,k^*}({\bf h}_{F,k^*})$ and $\beta_{M,k^*}({\bf h}_{M,k^*})$ are the path-loss (small-scale fading component) from the FAP and MBS to the selected femto-MT with index $k^*$, while $\alpha_{F,k^*}({\bf g}_{F,k^*})$ and $\alpha_{M,k^*}({\bf g}_{M,k^*})$ are the path-loss (small-scale fading component) from the FAP and MBS to the selected macro-MT with index $k^*$. The entries of all small-scale fading components are modeled as independent and identically distributed (i.i.d.) complex Gaussian random variables with zero mean and unit variance. Still in (\ref{y}), $P_F$ and $P_M$ denote the transmit power for FAP and MBS, respectively, and $n_F$ and $n_M$ are additive Gaussian noise with variance $\sigma^2_F$ and $\sigma^2_M$, respectively.
\vspace{-4mm}
\section{Low-Complexity QoS-Aware Coordinated Scheduling (QACS)}
\label{bdss}
\begin{figure*}
\centering
  \includegraphics[width=7in]{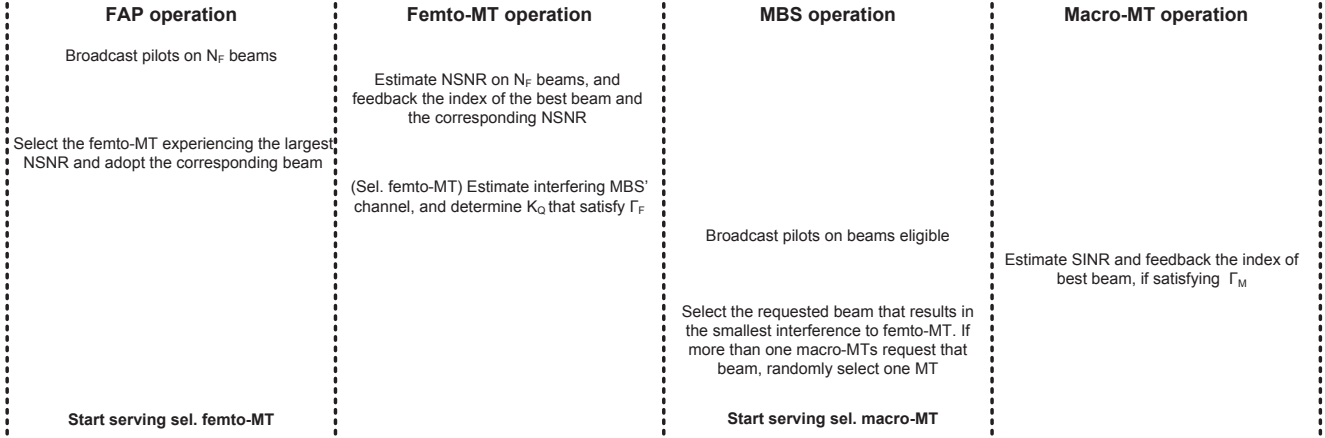}\vspace*{-4mm}\\
  \caption{{Mode of operations of QACS.}}\label{operation}
  \end{figure*}

In this section, we present the mode of operations of the proposed coordinated scheduling strategy, namely QoS-Aware Coordinated Scheduling (QACS) for a typical HetNet with one MBS and one FAP. The QACS mitigates the cross-tier interference through sequential beamforming design and user selection with limited information exchange between the FAP and the MBS. In particular, the FAP needs only to share the index of those qualified beams that the MBS can use. The detailed operation of the QACS {(depicted in Fig.~\ref{operation})} is summarized as follows.

a) The considered FAP starts its MT and beam selection by transmitting pilots on all $N_F$ beams to femto-MTs in its coverage. All femto-MTs then estimate their received normalized signal to noise ratio (NSNR) on different beams, which is proportional to $|{\bf h}_{F,k}^H {\bf w}_i|^2,1 \leq k \leq K_{F}, 1 \leq i \leq N_F$, and then feedbacks the maximum NSNR on all beams together with the index of the beam that achieves the maximum NSNR.

b) The FAP selects the MT achieving the largest NSNR among all femto-MTs, i.e., femto-MT $k^*$, where $\{i^*,k^*\}=\arg \max_{i,k}|{\bf h}_{F,k}^H {\bf w}_i|^2$, and adopts the corresponding beam ${\bf w}_{i^*}$ for transmission. We note that long-term fairness is guaranteed among all femto-MTs, as the MT selection policy depends only on the normalized channel statistics.

c) Femto-MT $k^*$ then estimates the channel from the interfering MBS, denoted by ${\bf h}_{M,k}$ \cite{MIMOest}. Based on the estimation of ${\bf h}_{M,k}$ as well as the knowledge of MBS's codebook (which is predefined), femto-MT $k^*$ is able to determine those MBS beams that lead to acceptable received signal to interference-plus-noise ratio (SINR). Without loss of generality, we suppose there are $N_Q \leq N_F$ qualified beams from MBS's codebook, such that
    \begin{equation}\label{tildegamma}
    \widetilde{\gamma}_{F,k^*}=\frac{P_F \beta_{F,k^*} |{\bf h}_{F,k^*}^H {\bf w}_{i^*}|^2}{P_M \beta_{M,k^*}|{\bf h}_{M,k^*}^H {\bf f}_{j}|^2+\sigma_F^2}\geq \Gamma_F,
    \end{equation}
for $1 \leq j \leq N_Q$. The indexes of these qualifying beams will also be fedback to the FAP, which are then sequentially forwarded to the MBS via the backhaul, based upon the value of the resulting $\widetilde{\gamma}_{F,k^*}$.

d) The MBS then begins its MT and beam selection, by transmitting pilots on those qualified beams to macro-MTs. Each macro-MT then determines its instantaneous received SINR on each available beam. In particular, the SINR of macro-MT $k$ on beam $j$ is obtained as
        \begin{equation}
        \gamma_{M,k,j}=\frac{{P_M} \alpha_{M,k} |{\bf g}_{M,k}^H {\bf f}_j|^2}{P_F \alpha_{F,k} |{\bf g}_{F,k}^H {\bf w}_{i^*}|^2+\sigma^2_M},
        \end{equation}
        for $1\leq k \leq K_M,1 \leq j \leq N_Q$. Each macro-MT then chooses its own best beam which achieves the largest SINR among all beams, namely, $j^*=\arg \max_{j}(\gamma_{M,k,j})$. The corresponding SINR $\gamma_{M,k}=\gamma_{M,k,j^*}$ is called the best beam SINR of macro-MT $k$. Each macro-MT then compares its best beam SINR with its QoS requirement, denoted by $\Gamma_M$. When its best beam SINR is greater than $\Gamma_M$, the macro-MT will feedback its best beam index {to request service on that beam}.

e) The MBS then selects the beam among the best beams of all qualified macro-MTs, such that the selected beam results in the smallest cross-tier interference to the selected femto-cell MT among all requested best beams. If only one macro-MT {{requests this beam}}, it will be the selected MT. If more than one {{requests it}}, the MBS randomly selects one of them in order to reduce the complexity. For the scheduling fairness, as the selected beam for macrocell is independent of the selected macro-MT, we argue that all macro-MTs have equal chance to be selected, as long as their reported best beam SINR is above the threshold $\Gamma_M$.

{We note that the proposed QACS scheme has low-complexity in the following aspects: 1) No data or CSI, but only several scalars, i.e., index in the predetermined codebook, are required to be shared between cells. 2) All MTs estimate and feedback are several scalars, while only the selected femto-MT needs to estimate the interference channel, which is a vector, during the overall scheduling procedure. 3) The proposed QACS schemes is based on random beamforming, where no CSI feedback from MTs is needed.}

According to the aforementioned mode of operations, the proposed QACS gives priority to femtocell, as indoor subscribers usually require services with much higher speed \cite{Ding}. If the MBS cannot suppress its cross-tier interference to the femto-MT at an acceptable level, it switches to another spectrum band to serve its macro-MTs. On the other hand, the QACS is general enough to apply in any two cells employing coordinated beamforming, by treating one of them as master cell (femtocell in this paper), while the other as slave cell (macrocell in this paper). {Moreover, if the fairness between macrocell and femtocell are needed, the MBS and FAP can take turns to initiate the QACS.}

{{In Section IV}}, we will carry out exact throughput analysis of the HetNet under the proposed QACS. The ergodic capacity of the femtocell and the macrocell are in general calculated as
\begin{equation}
\label{rate}
R_F=\int_{0}^{\infty}r(\gamma) f_{\gamma_F}(\gamma) d\gamma,\quad R_M=\int_{0}^{\infty}r(\gamma) f_{\gamma_M}(\gamma) d\gamma,
\end{equation}
with $r(\gamma)=\log_2(1+\gamma)$, where $f_{\gamma_{F}}(\gamma)$ and $f_{\gamma_{M}}(\gamma)$ represent respectively the probability density function (PDF) of the
selected femto-MT's and macro-MT's SINR, which will be derived in the following sections.
%With the QACS, the SINRs of the selected users in femtocell and macrocell are given by
%\begin{equation}
%\label{SINRR}
%\gamma_{F} = \frac{P_F \beta_{F,k^*} |{\bf h}_{F,k^*}^H {\bf w}_{i^*}|^2} {{P_M} \beta_{M,k^*} |{\bf h}_{M,k^*}^H {\bf
%f}_{j^*}|^2 + \sigma^2_F},
%\end{equation}
%and
%\begin{equation}
%\label{SINRMk}
%\gamma_{M}=\frac{{P_M} \alpha_{M,k} |{\bf g}_{M,k}^H {\bf f}_{j^*}|^2}{P_F \alpha_{F,k} |{\bf g}_{F,k}^H {\bf w}_{i^*}|^2+\sigma^2_M},,1\leq k \leq K_M, 1 \leq j \leq ,
%\end{equation}
%respectively.
\vspace{-3mm}
\section{Throughput Analysis of Two-Tier HetNet under QACS}\label{s3}
In this section, we analytically quantify the throughput of the two-tier HetNet considered in Section \ref{s2} for the proposed QACS.
\vspace{-8mm}
\subsection{Distribution of the Number of Qualified Beams $N_Q$ Guaranteeing Femto-MT's QoS}
With the proposed QACS, the number of qualified beams that satisfy the femto-MT's QoS requirement, is varying with the fading channel condition. In particular, the selected femto-MT will feedback  the index of those beams of the MBS that, if used in macro-cell transmission, will result in an acceptable SINR value. Mathematical speaking, the beam $j$ belonging to the MBS's codebook will be qualified if the corresponding SINR of the selected macro-MT, given by
\begin{equation}
\label{gj}
 \gamma_{F,j} =\frac{P_F \beta_{F,k^*}|{\bf h}_{F,k^*}^H {\bf w}_{i^*}|^2}
{P_M \beta_{M,k^*}|{\bf h}_{M,k^*}^H {\bf f}_{j}|^2  + \sigma^2_F}=\frac{x_F}{\lambda_F y_{F,j}+\mu_F}
\end{equation}
is greater than a fixed threshold, denoted by $\Gamma_F$, which is pre-determined based on a certain QoS requirement, for $1 \leq j \leq N_M$. As a result, the number of qualified beams in macro-cell will be random, depending on both ${\bf h}_{F,k^*}$ and ${\bf h}_{M,k^*}$. In the following, we first derive the probability mass function (PMF) of the number of qualified beams $N_Q$ that satisfies the selected Femto-MT's QoS.

As $|{\bf h}_{M,k^*}^H {\bf f}_{j}|^2$ is a standard
chi-square r.v. with two degrees of freedom
\cite{Hassibi}, the
probability that $\gamma_{F,j}$ given in (\ref{gj}) is greater than $\Gamma_{F}$ conditioning on $x_F$ can be calculated as
\begin{equation}
\label{gls}
\Pr(\gamma_{F,j} \geq
\Gamma_F\big{|}x_F)=\int^{\frac{x_F}{\Gamma_F}-\mu_F}_{0}
\frac{1}{\lambda_F} e^{-y/\lambda_F} dy= 1-e^{\frac{-x_F+\mu_F
\Gamma_F}{\Gamma_F \lambda_F}}.
\end{equation}
Since $\gamma_{F,j}$ are independent of each other, the probability that the number of qualified beams $N_Q$ is equal to $n$ given $x_F$ is given by \begin{eqnarray}
\label{ns}
\nonumber {{\Pr(N_Q=m\big{|}x_F)=}}&&\binom{N_M}{m} \left(1-e^{\frac{-x_F+\mu_F
\Gamma_F}{\Gamma_F \lambda_F}}\right)^{m} \cdot\\
&&\left(e^{\frac{-x_F+\mu_F
\Gamma_F}{\Gamma_F \lambda_F}}\right)^{N-m}.
\end{eqnarray}
Finally, after unconditioning with the PDF of $x_F$ given in (\ref{xF}) and carrying out the integration, we obtain
\begin{eqnarray}
\label{pr}
\nonumber &&{{\Pr(N_Q=m)=}}K_F N_F \binom{N_M}{m} \sum_{i=0}^m \binom{m}{i} \sum_{j=0}^{K_F N_F-1}  \\
\nonumber &&\binom{K_F N_F-1}{j}(-1)^{K_F N_F-1-j+m-i}\frac{e^{\frac{\mu_F}{\lambda_F}(N_M-i)}}{\frac{N_M-i}{\lambda_F \Gamma_F}+K_F N_F-j}.\\
\end{eqnarray}
\vspace{-6mm}
\subsection{Distribution of the MT and Beam selected by FAP}
Based on the mode of operation of QACS, the SINR of the selected femto-MT specializes to
\begin{equation}
\label{gammaR}
\gamma_{F} =  \frac{P_F \beta_{F,k^*} \max_{i,k} |{\bf h}_{F,k}^H {\bf
w}_{i}|^2}{P_M \beta_{M,k^*} y_F+\sigma^2_F}=\frac{x_F}{\lambda_F y_F+\mu_F},
\end{equation}
where $\gamma_F \geq \Gamma_F$ is required to be guaranteed. In order to obtain the distribution of $\gamma_F$, we need to consider $\gamma_{F,j}$ given in (\ref{gj}). In (\ref{gj}), $x_F$ is the random variable (r.v.) denoting the largest NSNR throughout all $N_F$ beams among all $K_F$ femto-MTs, which mathematically turns out to be the largest one of $N_F K_F$ i.i.d. random variables (r.v.s), as all beams are ortho-normal \cite{Hassibi} and different MTs experience independent channels.

%Its PDF is given by
%\begin{equation}\label{xF}
%f_{x_F}(x)=K_F N_F (1-e^{-x})^{K_F N_F-1} e^{-x},
%\end{equation}
%while $f_{\lambda_F y_{F,j}}=\frac{1}{\lambda_F} e^{-\frac{x}{\lambda_F}}$. By combining them together, the PDF and cumulative distribution function (CDF) of the fractional form is obtained as
%\begin{equation}
%\label{f1}
%f(x)=K_F N_F \sum_{i=0}^{K_F N_F-1}
%\binom{K_F N_F-1}{i} (-1)^{K_F N_F-1-i} \frac{e^{-(K_F N_F-i)\mu_{F} x}}{\lambda_{F}} \left( \frac{1+\mu_{F}
%[(K_F N_F-i)x+\frac{1}{\lambda_{F}}]}{[(K_F N_F-i)x+\frac{1}{\lambda_{F}}]^2}\right).
%\end{equation}
%\begin{equation}
%\label{F}
%F(x)=K_F N_F \sum_{i=0}^{K_F N_F-1}
%\binom{K_F N_F-1}{i} (-1)^{K_F N_F-1-i}
%\frac{\mu_{F}}{(K_F N_F-i)\lambda_{F}}\frac{e^{-(K_F N_F-i)\mu_{F}
%x}}{-(K_F N_F-i)\mu_{F} x + \frac{\mu_{F}}{\lambda_{F}}}.
%\end{equation}
%Based on $f(x)$ and $F(x)$, we can calculate the CDF and PDF of $\gamma_{F,j}$ as
%\begin{equation}
%F_{\gamma_{F,j}}(x)=\frac{\Pr(\gamma_{F,j} \leq x, \gamma_{F,j} \geq \Gamma_F)}{\Pr(\gamma_{F,j} \geq \Gamma_F)}=\frac{F(x)-F(\Gamma_F)}{1-F(\Gamma_F)}, ~f_{\gamma_{F,j}}(x)=\frac{f(x)}{1-F(\Gamma_M)},x \geq \Gamma_F.
%\end{equation}
%Then, the PDF of $\gamma_F$ can be written in terms of the PDF and CDF of $\gamma_{F,j}$ given the number of best beams, $N_B$, as
%\begin{equation}\label{pdfg1a}
%f_{\gamma_F}(x \big|N_B=n)=\sum_{j=1}^n \left(\prod_{l=1,l \neq j}^{n} F_{\gamma_{F,j}}(x)\right) f_{\gamma_{F,j}}(x).
%\end{equation}
%
On the other hand, according to the mode of operations described in Section \ref{s3}, $\lambda_F y_F$ represents the smallest projection  power from the channel vector ${\bf h}_{M,k^*}$ onto the MBS's best beam, $|{\bf h}_{M,k^*}^H {\bf f}_{j}|^2$, $1 \leq j \leq N_B$ among all $N_B$ best beams. It can be shown that $ |{\bf h}_{M,k^*}^H {\bf f}_{j}|^2,1 \leq j \leq N_B$ are truncated chi-square r.v.s with two degrees of freedom \cite{Hassibi}. The PDF of $\lambda_F y_{F,j}$ given $x_F$ is given by
\begin{equation}
f_{\lambda_F y_{F,j}}(y \big|x_F)=\frac{\frac{1}{\lambda_F}e^{-\frac{y}{\lambda_F}}}{\Pr(\gamma_{F,j} \geq \Gamma_F \big|x_F)}=\frac{\frac{1}{\lambda_F}e^{-\frac{y}{\lambda_F}}}{1-e^{\frac{-x_F+\mu_F
\Gamma_F}{\Gamma_F \lambda_F}}}.
\end{equation}
%Before analyzing $x_F$ and $\lambda_F y_F$, we first review an important result on ordered statistics. Given $\{\alpha_i\}^{N}_{i=1}$ with $\alpha_i,1 \leq i \leq N$ a r.v. with $F_\alpha(x)$ and $f_\alpha(x)$ as its cumulatative distribution function (CDF) and PDF, respectively, the PDF of the $l$th ($1 \leq l \leq N$) largest one among $N$, denoted by $\alpha_{l:N}$, is verified to be \cite{yangbook}
%\begin{equation}
%\label{lnk}
%f_{\alpha_{l:N}}(x)=\frac{(N!)}{(N-l)!(l-1)!} F_\alpha(x)^{N-l}  f_\alpha(x).
%\end{equation}
We then obtain the PDF of $x_F$ as well as $\lambda_F y_F$ conditioning on $x_F$ and the number of best beams $N_B$, as \cite{yang}
\begin{subequations}
\begin{equation}\label{xF}
f_{x_F}(x)=K_F N_F (1-e^{-x})^{K_F N_F-1} e^{-x},
\end{equation}
\begin{equation}\label{yF}
f_{\lambda_F y_F}(y\big{|}x_F,N_B=n)=\frac{\frac{n}{\lambda_F} e^{-\frac{n y}{\lambda_F}}}{1-e^{\frac{-x_F+\mu_F
\Gamma_F}{\Gamma_F \lambda_F}}},
\end{equation}
\end{subequations}
for $1 \leq n \leq N_B$, respectively. When $n=0$, i.e., no beams is qualified, MBS is turned off at this spectrum band and the selected femto-MT suffers from no cross-tier interference.

By combining $f_{x_F}(x)$ and $f_{\lambda_F y_F}(y\big{|}x_F,N_B=n)$, the PDF of $\gamma_F$ given $N_B=n$ is written as
\begin{eqnarray}\label{pdfg1a}
\nonumber {{f_{\gamma_F}(\gamma \big|N_B=n)=}}&&\int_0^{\infty} f_{\lambda_F y_F}(x/\gamma-\mu_F\big{|}x,N_B=n) \cdot\\
&& -x/\gamma^2 f_{x_F}(x) dx
\end{eqnarray}
%\begin{equation}
%f_{\gamma_F}(x\big{|}N_B=n)=K_F N_F \int^{\infty}_0
%(z+\mu_F) (1-e^{-x(z+\mu_F)})^{K_F N_F-1} e^{-x(z+\mu_F)} f_{\lambda_F y_F}(z) dz.
%\end{equation}
%After carrying out the integrations and some manipulations, we obtain the following closed-form expression: $f_{\gamma_F}(\gamma \big|N_B=n)=$
%\begin{equation}
%\label{pdfg1a}
%\frac{K_F N_F n}{\lambda_k} \sum_{i=0}^{K_F N_F-1} \binom{K_F N_F-1}{i} (-1)^{K_F N_F-1-i}e^{-\mu_F (K_F N_F-i)x}\left(\frac{\mu_F\left((K_F N_F-i)x+\frac{n}{\lambda_F}\right)+1}{\left((K_F N_F-i)x+\frac{n}{\lambda_F}\right)^2}\right),
%\end{equation}
for $1 \leq n \leq N_Q$, which is numerically verified to converge using mathematical software such as Maple and Mathematica. We then have $
f_{\gamma_F}(\gamma|N_B=0)=K_F N_F (1-e^{-\frac{\gamma}{\mu_F}})^{K_F N_F-1} e^{-\frac{\gamma}{\mu_F}}$ for $n=0$.
\vspace{-6mm}
\subsection{Distribution of the Macro-MT's SINR}
The SINR of the selected macro-MT under the QACS can be written as
\begin{equation}
\label{gammaM}
\gamma_{M}= \frac{{P_M} \alpha_{M,k^*} \max_{j}|{\bf g}_{M,k^*}^H {\bf f}_{j}|^2}{P_F \alpha_{F,k^*} |{\bf g}_{F,k^*}^H {\bf w}_{i^*}|^2+\sigma^2_M},1\leq j \leq N_Q,
\end{equation}
where $\gamma_M \geq \Gamma_M$ is required to be guaranteed. In order to obtain the distribution of $\gamma_M$, we need the following SINR expression
\begin{equation}
\gamma_{M,k}=\frac{{P_M} \alpha_{M,k} \max_{j}|{\bf g}_{M,k}^H {\bf f}_{j}|^2}{P_F \alpha_{F,k} |{\bf g}_{F,k}^H {\bf w}_{i^*}|^2+\sigma^2_M}=\frac{x_{M,k}}{\lambda_{M,k} y_{M,k}+\mu_{M,k}},
\end{equation}
for $1\leq k \leq K_M$, which denotes the best beam SINR reported at macro-MT $k$. We note that $x_{M,k}$ denotes the largest NSNR among $N_Q$ qualified beams, which PDF is given by
\begin{equation}
f_{x_{M,k}}(x)=N_Q (1-e^{-x})^{N_Q-1} e^{-x},
\end{equation}
while $f_{\lambda_{M,k} y_{M,k}}(y)=\frac{1}{\lambda_{M,k}}e^{-\frac{y}{\lambda_{M,k}}}$. By combining them together, the PDF and cumulative distribution function (CDF) of $\gamma_{M,k}$ can be obtained as
\begin{subequations}
\begin{eqnarray}
\label{f}
\nonumber f_{\gamma_{M,k}}(\gamma)&=& N_Q \sum_{i=0}^{N_Q-1}
\binom{N_Q-1}{i}  \frac{e^{-(N_Q-i)\mu_{M,k} \gamma}}{\lambda_{M,k}} \\
&& \nonumber (-1)^{N_Q-1-i}\left( \frac{1+\mu_{M,k}
[(N_Q-i)\gamma+\frac{1}{\lambda_{M,k}}]}{[(N_Q-i)\gamma+\frac{1}{\lambda_{M,k}}]^2}\right)\\
\end{eqnarray}
\begin{eqnarray}
\label{F}
\nonumber F_{\gamma_{M,k}}(\gamma)&=&N_Q \sum_{i=0}^{N_Q-1}
\binom{N_Q-1}{i}
\frac{\mu_{M,k}}{(N_Q-i)\lambda_{M,k}}\\
&& (-1)^{N_Q-1-i}\frac{e^{-(N_Q-i)\mu_{M,k}
\gamma}}{-(N_Q-i)\mu_{M,k} \gamma + \frac{\mu_{M,k}}{\lambda_{M,k}}},
\end{eqnarray}
\end{subequations}
%respectively.
Based on $f_{\gamma_{M,k}}(\gamma)$ and $F_{\gamma_{M,k}}(\gamma)$, we can calculate the CDF and PDF of $\gamma_M$ conditioning on the number of qualified beams, $N_Q$, given $K_Q \neq 0$, as
\begin{subequations}
\begin{equation}
F_{\gamma_M}(\gamma\big| N_Q=m)=\frac{F_{\gamma_{M,k}}(\gamma)-F_{\gamma_{M,k}}(\Gamma_M)}{1-F_{\gamma_{M,k}}(\Gamma_M)}
\end{equation}
\begin{equation}\label{fMnqm}
f_{\gamma_M}(\gamma\big| N_Q=m)=\frac{f_{\gamma_{M,k}}(\gamma)}{1-F_{\gamma_{M,k}}(\Gamma_M)},\gamma \geq \Gamma_M, K_Q \neq 0
\end{equation}
\end{subequations}
respectively. For $K_Q=0$, we simply have $F_{\gamma_M},f_{\gamma_M}(x\big| K_Q=0)=0$, since no MT is served in the macro-cell. Moreover, the probability that $K_Q \neq 0$ is obtained as
\begin{eqnarray}\label{prkq0nqm}
\nonumber &&{{\Pr(K_Q \neq 0\big|N_Q=m)}}=\\
\nonumber &&\sum_{k=1}^{K_M}\binom{K_M}{k} \left(F_{\gamma_{M,k}}(\Gamma_M\big|N_Q=m)\right)^{K_M-k} \\ &&\left(1-F_{\gamma_{M,k}}(\Gamma_M\big|N_Q=m)\right)^k.
\end{eqnarray}
\vspace{-10mm}
\subsection{Distribution of the Number of Best Beams $N_B$ Requested by Qualified macro-MTs}
Before performing the resulting throughput analysis, we also need the distribution of the number of best beams $N_B \leq N_Q$ requested by $K_Q$ qualified macro-MTs. This term is useful because the selected beam results in the smallest cross-tier interference to the selected femto-cell MT, which has effect on the femto-MT's resulting throughput. In particular, For either $N_Q=0$ or $K_Q=0$, we simply have $N_B=0$. For the more general case, i.e., $N_Q,K_Q \neq 0$, the probability that exactly $N_B$ best beams are active, given that $N_Q$ beams are qualified, is obtained as \cite[(15)]{yang}
%\begin{equation}
%\Pr(N_B=n\big|N_Q=m,K_Q=k)=\frac{1}{m^k} \binom{m}{n} \sum_{i=1}^n \binom{n}{i}(-1)^{n-i} i^k,1 \leq n \leq m.
%\end{equation}
\begin{eqnarray}
\nonumber &&\Pr(N_B=n\big|N_Q=m)=\binom{m}{n} \sum_{i=1}^n \binom{n}{i}(-1)^{n+i} \\
&& \bigg[\left(\frac{i}{m} +\frac{m-i}{m} F_{\gamma_{M,k}}(\Gamma_M)\right)^{K_M}-(F_{\gamma_{M,k}}(\Gamma_M))^{K_M}\bigg].
\end{eqnarray}
By removing the condition, we have $\Pr(N_B=n)=\sum_{m=1}^{N_M} \Pr(N_Q=m) \Pr(N_B=n\big|N_Q=m)$, with $\Pr(N_Q=m)$ given in (\ref{pr}).
\vspace{-6mm}
\subsection{Throughput Analysis}
Finally, by combining (\ref{pdfg1a}), (\ref{prkq0nqm}), (\ref{pr}), and (\ref{fMnqm}) into (\ref{rate}), the throughputs of the femtocell and the macrocell under the QACS can be calculated as
\begin{subequations}
\begin{equation}
\label{RR}
R_F=\sum_{n=0}^{N_B} \Pr(N_B=n) \int_0^{\infty} r(\gamma)f_{\gamma_F}(\gamma \big{|}N_B=n)d \gamma
\end{equation}
\begin{equation}
\label{RM}
R_M=\sum_{m=1}^{N_M} \Pr(K_Q \neq 0,N_Q=m)\int_0^{\infty}r(\gamma)f_{\gamma_M}(\gamma \big{|}N_Q=m)d \gamma,
\end{equation}
\end{subequations}
respectively. Final expressions, which are omitted here for brevity, can readily be numerically evaluated using Maple and Mathematica.

%We note that $\Pr(N_b=0\big|K_Q=k,N_Q=m)$ in (\ref{RR}) disappears, as the probability is zero. This is because when $k>0$, at least one best beam satisfying the macro-MT's QoS requirement is avialable at the MBS. If either $N_Q$ or $K_Q$ is zero, this impies that no beam in MBS's codebook can satisfy the QoS requirement of the selected femto-MT, or none of the macro-MTs' best beam SINR meets its QoS requirement. In either case, the transmission of macro-cell is shut off and the selcted femto-MT suffers from no interference. To this end, its corresponding received SNR reduces to $\gamma_F=\frac{x_F}{\mu_F}$, whose PDF is easily obtained as

%\begin{equation}
%R_{\rm MAC}=\int_u \int_v \log_2 \left(1+\frac{P_{\rm MAC} \cdot u}{P_{\rm FAP} \cdot v +\sigma^2}\right)\leq \log_2 \left(1+\int_u \int_v\frac{P_{\rm MAC} \cdot u}{P_{\rm FAP} \cdot v +\sigma^2}\right)=\overline{R}_{\rm MAC}.
%\end{equation}
%
%\begin{equation}
%\overline{R}_{\rm MAC}=\log_2 \left(1+\frac{P_{\rm MAC}}{P_{\rm FAP}}\overline{u} \int_v \frac{1}{v+\frac{\sigma^2_n}{P_{\rm FAP}}}\right) \approx \log_2 \left(1+\frac{P_{\rm MAC}}{P_{\rm FAP}} \frac{\overline{u}}{\overline{v}+\frac{\sigma^2_n}{P_{\rm FAP}}}\right).
%\end{equation}
\vspace{-4mm}
\section{Numerical Examples}
\begin{figure}
\centering
  \includegraphics[width=3.5in]{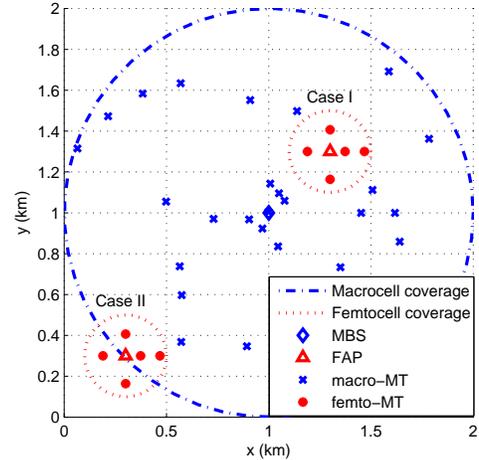}\vspace*{-4mm}\\
  \caption{Simulation scenario includes a macrocell with radius $1$km, and a femtocell with radius $20$m dropped in the macrocell. There are $K_M=50$ macro-MTs and $K_F=5$ femto-MTs uniformly distributed in their respective coverage. We have $P_M=50$dBm at the $N_M=4$-antenna MBS and $P_F=20$dBm at the $N_F=2$-antenna FAP. The path-loss model adopted in the rest of the paper follows that in \cite[Table II]{Ding}. Two cases are considered: (Case I) the distance between MBS and FAP is $100$m and (Case II) the distance between MBS and FAP is $800$m. }\label{sim}
  \end{figure}

\begin{figure}
\centering
\includegraphics[width=3.5in]{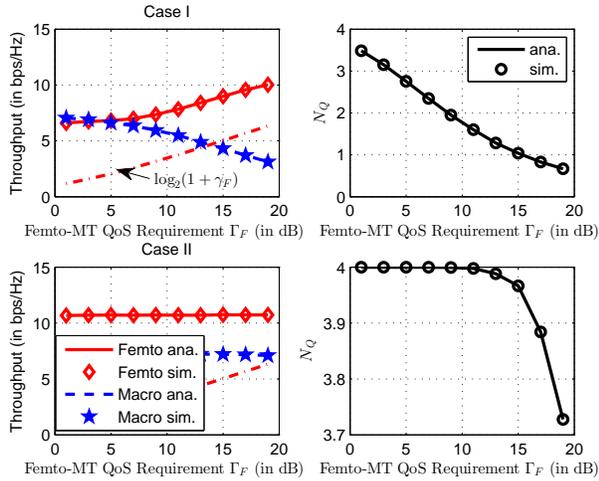}\vspace*{-4mm}\\
    \caption{Femtocell/Macrocell throughput versus femto-MT's QoS requirement $\Gamma_F$ for a system with $\Gamma_M=10$ dB.}\label{fig1}
\end{figure}

In this section, we present selected numerical examples to illustrate the mathematical formalism on the throughput analysis for the proposed QACS. The analytical results derived in the paper will all be verified through Monte-Carlo simulations. The simulation scenario is depicted in Fig.~\ref{sim}.
In Fig.~\ref{fig1}, we depict the femtocell/macrocell throughput as the function of the femto-MT's SINR requirement $\Gamma_F$. The number of qualified beams at the macrocell's codebook that satisfy the QoS requirement of the femto-MT's QoS, $N_Q$ is also provided in Fig.~\ref{fig1} for reference. According to the figure, for both cases, the femtocell's throughput is increasing in $\Gamma_F$, while the macro-cell's throughput as well as $N_Q$ are both decreasing. This is because when the QoS requirement of the femto-MT increases, the FAP becomes more strict when selecting the qualified beams for macrocell transmission, which improves its own throughput, while suppressing the macrocell's throughput by reducing the number of candidate beams for macrocell transmission. A further observation will be when the FAP get closer to the MBS, cf. Case I, the proposed QACS is more efficient to improve the femtocell throughput, at the expense of reducing macrocell's throughput. More precisely, in the region of $\Gamma_F > 15$dB, the macrocell's throughput suffers dramatically. In this case, the MBS may consider the communication on the other spectrum band with less cross-tier interference.

\begin{figure}
\centering
\includegraphics[width=3.5in]{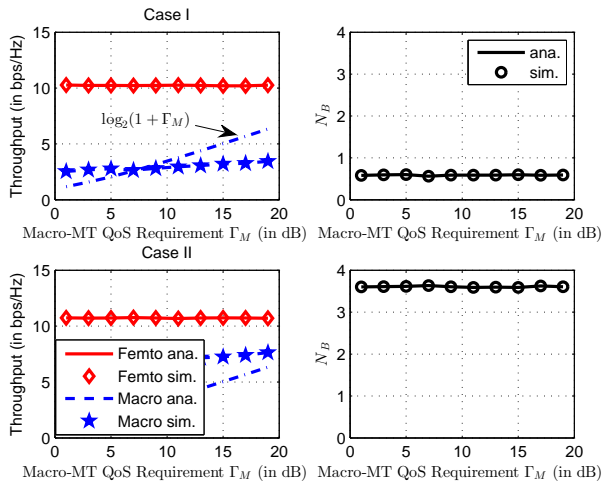}\vspace*{-4mm}\\
\caption{Femtocell/Macrocell throughput versus macro-MT's QoS requirement $\Gamma_M$ for a system with $\Gamma_F=20$ dB.}\label{fig2}
\end{figure}

Fig.~\ref{fig2} plots the femtocell/macrocell throughput and the number of best beams requested by macro-MTs, $N_B$ as the function of the macro-MT's SINR requirement $\Gamma_M$. The curves have manifested very different behaviors compared with those in Fig.~\ref{fig1}. In particular, both the femtocell throughput and $N_B$ keep unchanged covering a wide range of $\Gamma_M$, while the macrocell throughput slightly improves with the increasing $\Gamma_M$. By comparing Case I and II, we notice that with current parameters, macro-MTs are not always able to achieve their target QoS when $\Gamma_M > 10$dB in Case I, while the threshold has been relaxed to $\Gamma_M > 20$dB in Case II, as the the interference from the FAP is negligible.

\begin{figure}
\centering
\includegraphics[width=3.5in]{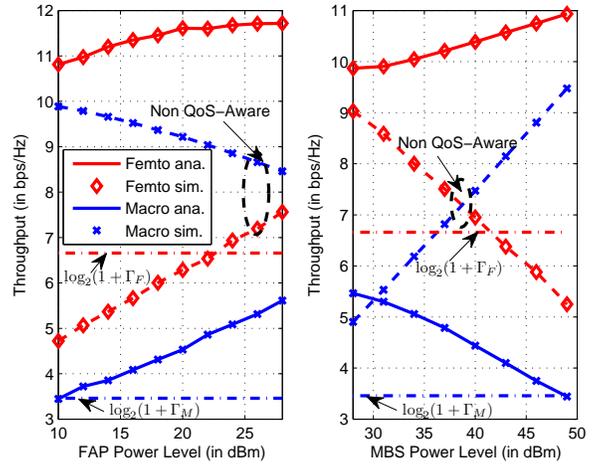}\vspace*{-4mm}\\
\caption{{Femtocell/Macrocell throughput versus transmit power level for a system with $\Gamma_F=20$dB, $\Gamma_M=10$dB, and the distance between MBS and FAP is 500m.}}\label{fig3}
\end{figure}

{{In Fig.~\ref{fig3}, we compare the femtocell/macrocell throughput achieved by the proposed QACS with traditional coordinated beamforming without QoS consideration, cf. \cite{icc_zhu}. The resulting throughputs are depicted as functions of transmit power level at FAP and MBS, respectively. We observe from the figure that the proposed QACS is able to satisfy QoS requirements for both femtocell and macrocell, for any transmit power choice at MBS and FAP, with the expense that the selected femto-MT needs to determine $K_Q$ that satisfies $\Gamma_F$, and each macro-MT needs to determine if its best beam achieved SINR satisfies $\Gamma_M$.}}


\vspace{-4mm}
\begin{thebibliography}{20}
\bibitem{hetnet}
J. Montojo, Y. Wei, T. Ji, T. Luo, M. Vajapeyam, T. Yoo, O. Song, and D. Malladi, ``A survey on 3GPP heterogeneous networks,'' \emph{IEEE Wireless Commun. Mag.}, vol. 18, no. 3, pp. 10-21, Jun. 2011.

\bibitem{femto1}
V. Chandrasekhar, J. Andrews, and A. Gatherer, ``Femtocell networks: a survey,'' \emph{IEEE Commun. Mag.}, vol. 46, no. 9, pp. 59-67, Sept. 2008.

\bibitem{femto2}
A. Barbieri, A. Damnjanovic, T. Ji, J. Montojo, Y. Wei, D. P. Malladi, O. Song, and G. Horn, ``LTE femtocells: System design and performance analysis,'' \emph{IEEE Journal Sel. Areas Commun.}, vol. 30, no. 3, pp. 586-594, Apr. 2012.

\bibitem{femto3}
J. G. Andrews, H. Claussen, M. Dohler, S. Rangan, and M. C. Reed, ``Femtocells: Past, present, and future,'' \emph{IEEE Journal Sel. Areas Commun.}, vol. 30, no. 3, pp. 497-508, Apr. 2012.

\bibitem{Ding}
A. R. Elsherif, Z. Ding, and X. Liu, ``Dynamic MIMO precoding for femtocell interference mitigation,'' \emph{IEEE Trans. Commun.}, vol. 62, no. 2, pp. 648-666, Feb. 2014.

\bibitem{interference}
T. Zahir, K. Arshad, A. Nakata, and K. Moessner, ``Interference management in femtocells,'' \emph{IEEE Commun. Surveys and Tutorials}, vol. 15, no. 1, pp. 293-311, First Quarter 2013.

\bibitem{icc_zhu}
J. Zhu and H. -C. Yang, ``Interference control with beamforming coordination for two-tier femtocell networks and its performance analysis,'' in \emph{Proc. IEEE Int. Conf. Commun. (ICC) 2011}, pp. 1-5, Jun. 2011.

\bibitem{dai}
Y. Dai, S. Jin, L. Pan, X. Gao, L. Jiang, and M. Lei, ``Interference control based on beamforming coordination for heterogeneous network with RRH deployment,'' \emph{IEEE System Journal}, vol. 9, no. 1, Mar. 2015.

\bibitem{mimohetnet}
H. S. Dhillon, M. Kountouris, and J. G. Andrews, ``Downlink MIMO HetNets: Modeling, Ordering results, and performance analysis,'' \emph{IEEE Trans. Wireless Commun.}, vol. 12, no. 10, pp. 5208-5222, Oct. 2013.

\bibitem{comp}
R. Irmer, \textit{et al.}, ``Coodinated multipoint: Concepts, performance, and field trial results,'' \emph{IEEE Commun. Mag.}, vol. 49, no. 2, pp. 102-111, Feb. 2011.

\bibitem{LG09}
LG Electronics, ``CoMP configurations and UE/eNB behaviors in LTE-Advanced,''
R1-090213, 3GPP TSG RAN WG1 Meeting \#55b, 2009.

\bibitem{huang}
S. He, Y. Huang, S. Jin, and L. Yang, ``Coodinated beamforming for energy efficient transmission in multicell multiuser systems,'' \emph{IEEE Trans. Commun.}, vol. 61, no. 12, pp. 4961-4971, Dec. 2013.

\bibitem{gc_zhu}
J. Zhu and H. -C. Yang, ``Low-complexity coordinated beamforming transmission for multiuser MISO systems and its performance analysis,'' in \emph{Proc. IEEE Global Commun. Conf. (Globecom) 2010}, pp. 1-5, Dec. 2010.

\bibitem{eurasip_zhu}
J. Zhu and H. -C. Yang, ``Performance analysis of low-complexity dual-cell random beamforming transmission with user scheduling,'' \emph{EURASIP Journal Wireless Commun. and Networking}, 2011: 191, pp. 1-11, Dec. 2011.

\bibitem{gc2_zhu}
J. Zhu, H. -C. Yang, and R. Schober, ``Performance evaluation of coordinated dual-cell transmission based on random unitary beamforming with user scheduling,'' in \emph{Proc. IEEE Global Commun. Conf. (Globecom) 2012}, pp. 1-5, Dec. 2012.

\bibitem{3GPP}
Samsung, ``Inter-Cell Interference Mitigation Through Limited Coordination,''
\emph{3GPP TSG RAN WG1}, Jeju, Korea, Aug. 2008.

\bibitem{LG09a}
LG Electronics, ``Codebook-based PMI restriction for LTE-Advanced
system,'' R1-090212, 3GPP TSG RAN WG1 Meeting \#55b, 2009.

\bibitem{hetnet5G}
M. Peng, Y. Li, Z. Zhao, and C. Wang, ``System architecture and key technologies for 5G heterogeneous cloud radio access networks,''\emph{IEEE Network}, vol. 29, no. 2, pp. 6-14, Mar. 2015.

\bibitem{Hassibi}
M. Sharif, and B. Hassibi, ``On the capacity of MIMO broadcast channels
with partial side information,'' \emph{IEEE Trans. Inform. Theory}, vol. 51, no.
2, pp. 506-522, Feb. 2005.

\bibitem{MIMOest}
{{M. Biguesh and A. B. Gershman, ``Training-based MIMO channel estimation: a study of estimator tradeoffs and optimal training signals,'' \emph{IEEE Trans. Sig. Proc.}, vol. 54, no. 3, pp. 884--893, Mar. 2006.}}

%\bibitem{yangbook}
%H.-C. Yang and M.-S. Alouini, \emph{Order Statistics in Wireless Communications},
%Cambridge University Press, 2011.

\bibitem{yang}
P. Lu and H. -C. Yang,``A simple and efficient user-scheduling strategy for RUB-based multiuser MIMO systems and its sum-rate analysis,'' \emph{IEEE Trans. Vehi. Tech.}, vol. 58, no. 9, Nov. 2009.
\end{thebibliography}
\end{document}